\documentclass[journal=jacsat,manuscript=article]{achemso}

\usepackage[version=3]{mhchem} 
\usepackage{graphicx}
\usepackage{placeins} 
\graphicspath{{./Images/}}
\usepackage{caption}
\usepackage{subcaption}
\usepackage{graphicx}
\usepackage{booktabs}

\usepackage{chemformula}
\usepackage{siunitx}
\usepackage{lmodern}

\DeclareSIUnit\angstrom{\text{\AA}}

\author{Jonas Hänseroth}
\affiliation{Department of Theoretical Solid State Physics, Institute of Physics, Technische Universität Ilmenau, 98693 Ilmenau, Germany}
\alsoaffiliation{Theoretical Chemistry, Martin-Luther Universität Halle-Wittenberg, 06120 Halle (Saale), Germany}

\author{Daniel Sebastiani}
\affiliation{Theoretical Chemistry, Martin-Luther Universität Halle-Wittenberg, 06120 Halle (Saale), Germany}

\author{Jakob Scholl}
\affiliation{Fraunhofer-Institut für Keramische Technologien und Systeme IKTS, 
 Hydrogen Technologies, 99310 Arnstadt, Germany}

\author{Karl Skadell}
\affiliation{Fraunhofer-Institut für Keramische Technologien und Systeme IKTS, 
 Hydrogen Technologies, 99310 Arnstadt, Germany}

\author{Christian Dreßler}
\affiliation{Department of Theoretical Solid State Physics, Institute of Physics, Technische Universität Ilmenau, 98693 Ilmenau, Germany}
\email{christian.dressler@tu-ilmenau.de}

\title{Hydroxide Mobility in Aqueous Systems: Combining \textit{Ab Initio} Accuracy with Millisecond Timescales}

\begin{document}

\begin{abstract} 

We present a multiscale simulation approach for hydroxide transport in aqueous solutions of potassium hydroxide, combining \textit{ab initio} molecular dynamics (AIMD) simulations with force field ensemble averaging and lattice Monte Carlo techniques. This method achieves near \textit{ab initio} accuracy by capturing the femtosecond scale dielectric relaxation dynamics of the aqueous hydrogen bonding network, while extending the simulation capability to millisecond diffusion timescales. This extraordinary extension of the available length and time scales enables future studies of hydroxide mobility in functional materials such as nanostructured anion-exchange membranes, where hydroxide ions migrate through nanometer-sized channels. Remarkably, our approach demonstrates that a single AIMD trajectory is sufficient to predict hydroxide conductivity over a range of concentrations, underscoring its computational efficiency and relevance to the design of advanced energy materials.

\end{abstract}

\FloatBarrier

\FloatBarrier

\section{Introduction}
Green hydrogen, produced by the electrolysis of water using renewable electricity, plays a crucial role in the global transition to a more sustainable energy economy.\cite{hren2021} Among the various methods for water splitting, anion-exchange membrane (AEM) water electrolysis stands out due to its high efficiency and its ability to utilize inexpensive and abundant electrode materials, such as iron and nickel.\cite{mukerjee2021, henkensmeier2020, park2019, leng2012, dekel2018} In contrast, proton exchange membrane (PEM) electrolysis operates under acidic conditions and relies on noble metal catalysts like platinum and iridium, which are scarce and costly.\cite{alia2021, zou2015, schalenbach2018}

Two key challenges for advancing AEM technology are improving membrane stability under alkaline operating conditions and enhancing hydroxide conductivity.\cite{henkensmeier2020, wijaya2024} Achieving the latter would benefit greatly from simulation tools capable of predicting hydroxide mobility with low computational effort. Such tools would enable the optimization of AEM materials before synthesis, accelerating their development. However, simulating the complex polymeric systems with solvated nanochannels, as found in AEMs, requires large-scale supercomputers and remains beyond the reach of state-of-the-art molecular dynamics (MD) simulations. Current simulations of hydroxide dynamics are often limited to small model systems, where the polymeric structure of the membrane is replaced by small organic molecules that mimic its functional groups.\cite{park2017, takaba2017, zelovich2024, zelovich2019mimics, zelovich2019hydration, zelovich2020, zelovich21_ohvsh3o}

While force field molecular dynamics (FFMD) simulations are computationally less demanding, they are inadequate for modeling hydroxide mobility.\cite{delucas2024} This is because hydroxide ion transport involves bond-breaking and bond-forming events, which can only be captured accurately using quantum chemical methods.\cite{zelovich2024, tuckerman2006acs, ouma2022}

To address this computational bottleneck, two alternative approaches have been proposed. The first one employs machine-learning-based interatomic force fields, which can simulate hydroxide ion mobility on timescales almost comparable to classical MD simulations while retaining near \textit{ab initio} accuracy.\cite{hellstrom2018, karibayev22_aemmlff, jinnouchi23_proton} The second approach combines MD simulations with other methods, such as Monte Carlo simulations, to model ion dynamics on much larger timescales, extending up to milliseconds.

In this article, we focus on the second approach, adapting our previously developed combined Molecular Dynamics/Lattice Monte Carlo (cMD/LMC) framework for simulating proton dynamics to hydroxide ions.\cite{kabbe2014, kabbe2016, kabbe2017, dressler2016} This adaptation is motivated by the mechanistic similarities and differences between the conduction of hydronium ions (\ce{H3O+}) and hydroxide ions (\ce{OH-}). While \ce{H3O+} forms via the addition of a proton to a water molecule, \ce{OH-} forms through proton removal. Both exhibit enhanced mobility compared to water molecules, attributed to the Grotthuss mechanism, which facilitates ion transport through a combination of hopping and reorientation steps.\cite{grotthuss, marx2006, tuckerman2006acs, tuckerman1995jpc}

The Grotthuss mechanism involves proton transfer along a chain of water molecules, propagating charge without significant molecular diffusion. The process includes a hopping step, where a proton is transferred along the chain, and a subsequent reorientation of water molecules to allow further transfers. Due to the Grotthuss mechanism, increased mobility can be observed for both ion types, with protons exhibiting greater mobility than hydroxide ions, as reflected in their diffusion coefficients (D(\ce{H+}) / D(\ce{H2O}) = 10 , D(\ce{OH-}) / D(\ce{H2O}) = 4).\cite{tuckerman2006acs} This difference arises from distinct conduction mechanisms between protons and hydroxide ions.

The "proton-hole mechanism" suggests a one-to-one correspondence between proton and hydroxide conduction, involving a threefold coordinated \ce{OH-} and an intermediate \ce{H3O2-} complex analogous to \ce{H3O+(H2O)3} and \ce{H5O2+}.\cite{zatsepina1972statingprotonhole} However, \textit{ab initio} studies by Tuckerman et al. demonstrate a distinct hydroxide transfer mechanism.\cite{tuckerman2002nature} The initial step involves a transition from a square-planar coordinated hydroxide ion (\ce{OH- (H2O)4}) to a tetrahedral geometry (\ce{OH- (H2O)3}). Notably, the Zundel-analog complex (\ce{H3O2-}) exists only transiently, persisting for just 2–3 oscillation periods during the transfer mechanism.\cite{roberts2009zundeltransientir} This behavior, observed through time-resolved IR experiments, aligns with the "presolvation concept", which emphasizes hypercoordination and dynamic solvation shell changes.\cite{tuckerman2006acs, tuckerman2010rev, smiechowski2007irdonatinghbond}

In this work, we evaluate our multiscale approach for simulating hydroxide transport in aqueous potassium hydroxide solutions. We demonstrate that a single \textit{ab initio} molecular dynamics trajectory within our framework is sufficient to predict hydroxide conductivity across a range of concentrations.

\FloatBarrier

\section{Method}
\begin{figure}[h!]
    \centering
    \begin{minipage}[t]{0.95\textwidth}
        \begin{tikzpicture}
            \centering
            \node[anchor=south west,inner sep=0] (image) 
            at (0,0) 
            {\includegraphics[width=\textwidth]{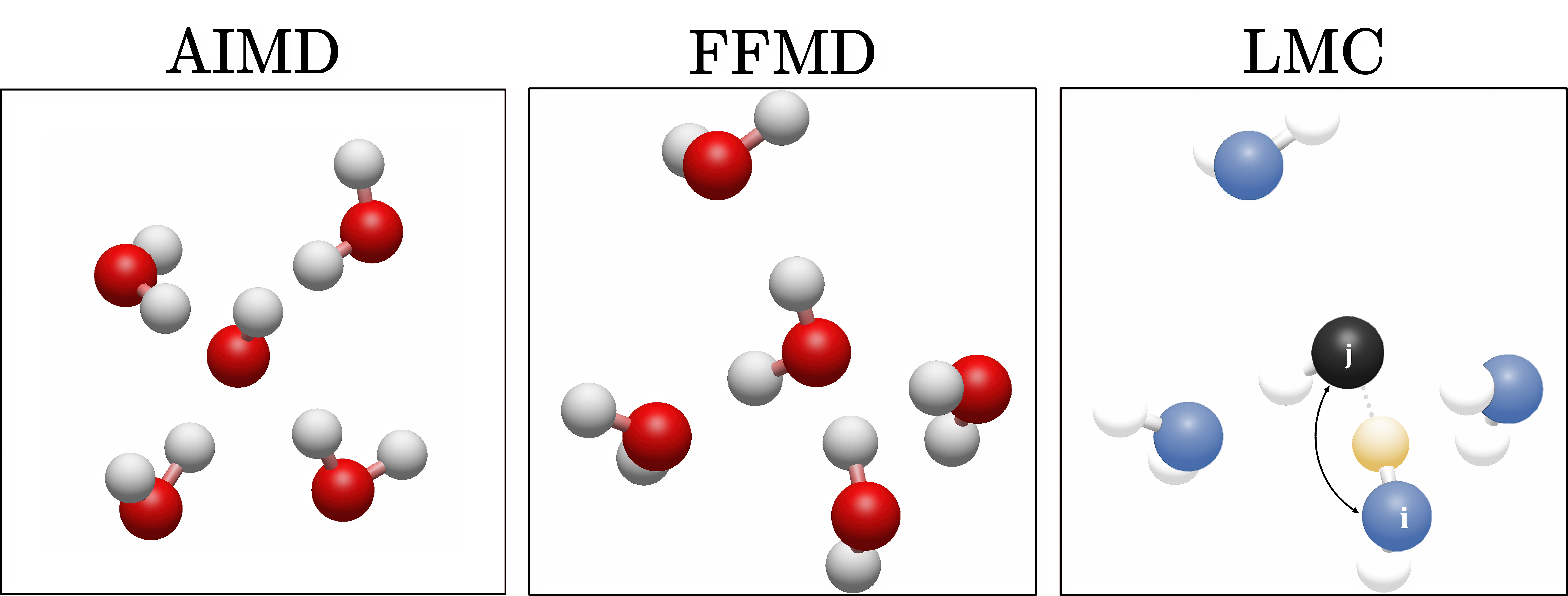}};
            \begin{scope}[x={(image.south east)},y={(image.north west)}, overlay]
                \node at  (0.81,0.19) {%
                    \begin{tabular}{@{}l@{}}
                        $\omega (d_{ij})$
                    \end{tabular}%
                };
            \end{scope}
        \end{tikzpicture}
    \end{minipage}
    \caption{Local hydroxide transfer rates derived from an AIMD simulation of aqueous KOH and an oxygen lattice derived from an FFMD simulation of pure water are required for the cMD/LMC approach. The long-range transport of hydroxide ions is calculated using these inputs in a Monte Carlo algorithm. The red atoms represent oxygen, while the white atoms correspond to hydrogen. Blue lattice sites denote water species, and the black lattice site represents a hydroxide species in the LMC approach. A proton undergoing a jump is highlighted in yellow, its probability of jumping from O\textsubscript{i} to O\textsubscript{j} is given by the term $\omega (d_{ij})$.}
    \label{img:overview}
\end{figure}

Our approach combines molecular dynamics simulations with a lattice Monte Carlo method to model hydroxide ion transport in aqueous solutions. The \textit{ab initio} molecular dynamics (AIMD) simulations capture local hydroxide transfer rates on the sub-picosecond timescale, while the Monte Carlo method simulates the long-range propagation of hydroxide ions on an oxygen lattice derived from a force field molecular dynamics simulations. Figure \ref{img:overview} illustrates the approach.

In the Monte Carlo approach, the system is simplified by representing only the oxygen atoms, which are either occupied by two protons (water molecules) or a single proton (hydroxide ions). The hydrogen atoms are fixed at the positions of their covalently bonded oxygen atoms, with the oxygen atom positions extracted from force field MD trajectories. While FFMD cannot simulate bond breaking or formation, our multiscale approach overcomes this limitation by enabling proton jumps between neighboring oxygen atoms. The probabilities of these proton jumps are determined using a jump rate function derived from quantum chemical simulations of solvated hydroxide ions. 

This Monte Carlo method uniquely combines standard lattice-based Monte Carlo and traditional kinetic Monte Carlo (KMC) approaches. In our method, the lattice is constructed using the oxygen positions from an underlying molecular dynamics trajectory, which are updated after each Monte Carlo step. Kinetic rates for proton transfer between neighboring lattice sites are periodically applied, with the fixed timestep determined by the time interval between consecutive frames in the MD trajectory.

\subsection{Sampling hydroxide jump rates from \textit{ab initio} molecular dynamics simulation}

\begin{figure}[h!] 
    \centering \includegraphics[width=1\textwidth]{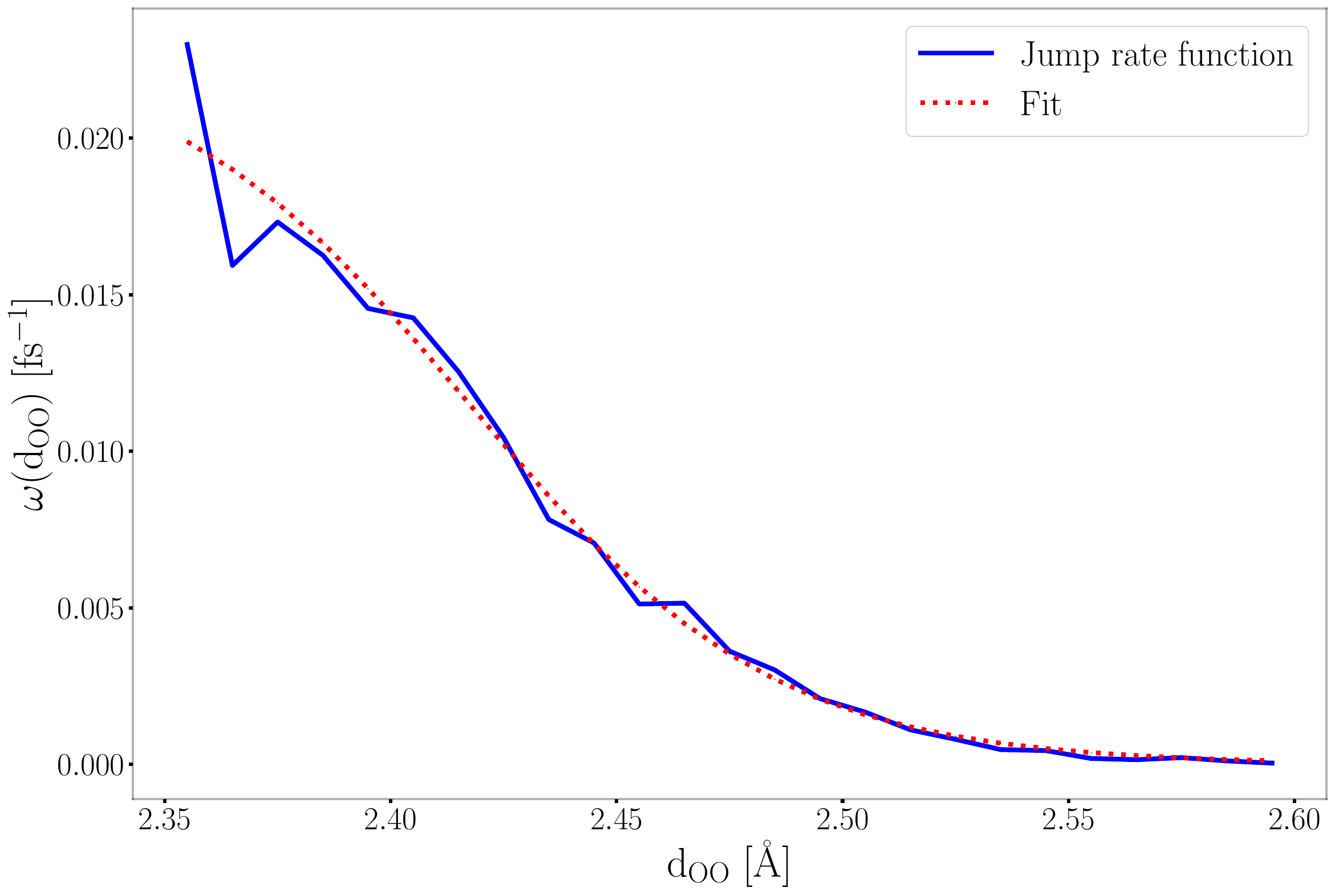} 
    \caption{Jump probability for protons with respect to the \ce{O-O} distance between the hydroxide ion and a neighboring water molecule. The jump rate function is sampled from the  AIMD trajectory of aqueous KOH solution $c(\mathrm{KOH})=$ \SI{17.89}{\mole\per\liter} at \SI{333}{\kelvin}.} 
    \label{img:jump_rates} 
\end{figure}

Previous studies have shown that proton transfer within the oxygen lattice is governed by the oxygen-oxygen distances.\cite{kabbe2014, kabbe2016, kabbe2017, dressler2016} It turns out that the relationship between proton transfer probabilities and oxygen-oxygen distances can be accurately described using a Fermi-like function (refer to Figure \ref{img:jump_rates} and Equation \ref{eq:fermi}).
The shape of this function is determined prior to the actual cMD/LMC runs on the basis of a comparably short \textit{ab initio} molecular dynamics simulation. From this MD simulation, we compute $\omega(d_{\mathrm{OO}})$ as the conditional probability for a jump at a given distance $d_{\mathrm{OO}}$ by counting the actual number of real jumps in the MD trajectory at this  distance divided by the number of the overall occurrence of this oxygen-oxygen distance between hydroxide ions and water molecules in the MD trajectory.

For an accurate description of the resulting rate function $\omega(d_{\mathrm{OO}})$ it is sufficient to have a modest number of proton jumps (per distance window) in the \textit{ab initio} trajectory; it is not necessary to perform an extended \textit{ab initio} simulation with a well-converged proton diffusion statistics.

The numerically obtained conditional hopping probability is fitted to a Fermi-Function according to Equation \ref{eq:fermi}. The fit parameters for our system (aq. KOH solution at $c=$ \SI{17.89}{\mole\per\liter} and at \SI{333}{\kelvin}) is given in Table \ref{tab:fermi_fit}.

\begin{equation} 
    \omega(d_{ij}) = \frac{a}{1 + \exp{\left(\frac{d_{ij}-b}{c}\right)}}
    \label{eq:fermi}
\end{equation}

\begin{table}[h!]
    \centering
    \caption{Fermi fit parameters describing the jump rate function of the aqueous KOH solution with $c(\mathrm{KOH})=$ \SI{17.89}{\mole\per\liter} at \SI{333}{\kelvin}.}
    \label{tab:fermi_fit}
    \begin{tabular}{l l l}
        \toprule
        $a$ [\si{\per\femto\second}]     &   $b$ [\si{\angstrom}]   & $c$ [\si{\per\angstrom}]  \\
        0.023 & 2.4 & 30 \\
        \bottomrule
    \end{tabular}
\end{table}

Naturally, the jump probability $\omega(d_{\mathrm{OO}})$ and thus its fit parameters ($a$, $b$, $c$) are concentration dependent. However we have explicitly checked that their variation within out concentration range is below \SI{10}{\percent} for any parameter and does not exhibit a systematic trend. Thus, there variations are of the same amplitude as the statistical fluctuations in the numerically computed conditional hopping probabilities. Therefore, we have chosen to work with a single set if parameters $a$, $b$, $c$ for all KOH concentrations.

\subsection{Lattice Monte Carlo Algorithm}

\begin{figure}[h!]
\vspace{0.5cm}
    \centering
    \begin{minipage}[t]{0.61\textwidth}
        \vspace{1cm}
        \begin{subfigure}[t]{\textwidth} 
        \vfill
            \centering
            \begin{tikzpicture}
    	 		\centering
    	 		\node[anchor=north west,inner sep=0] (image) 
    	 		at (0,0) 
    	 		{\includegraphics[width=\textwidth]{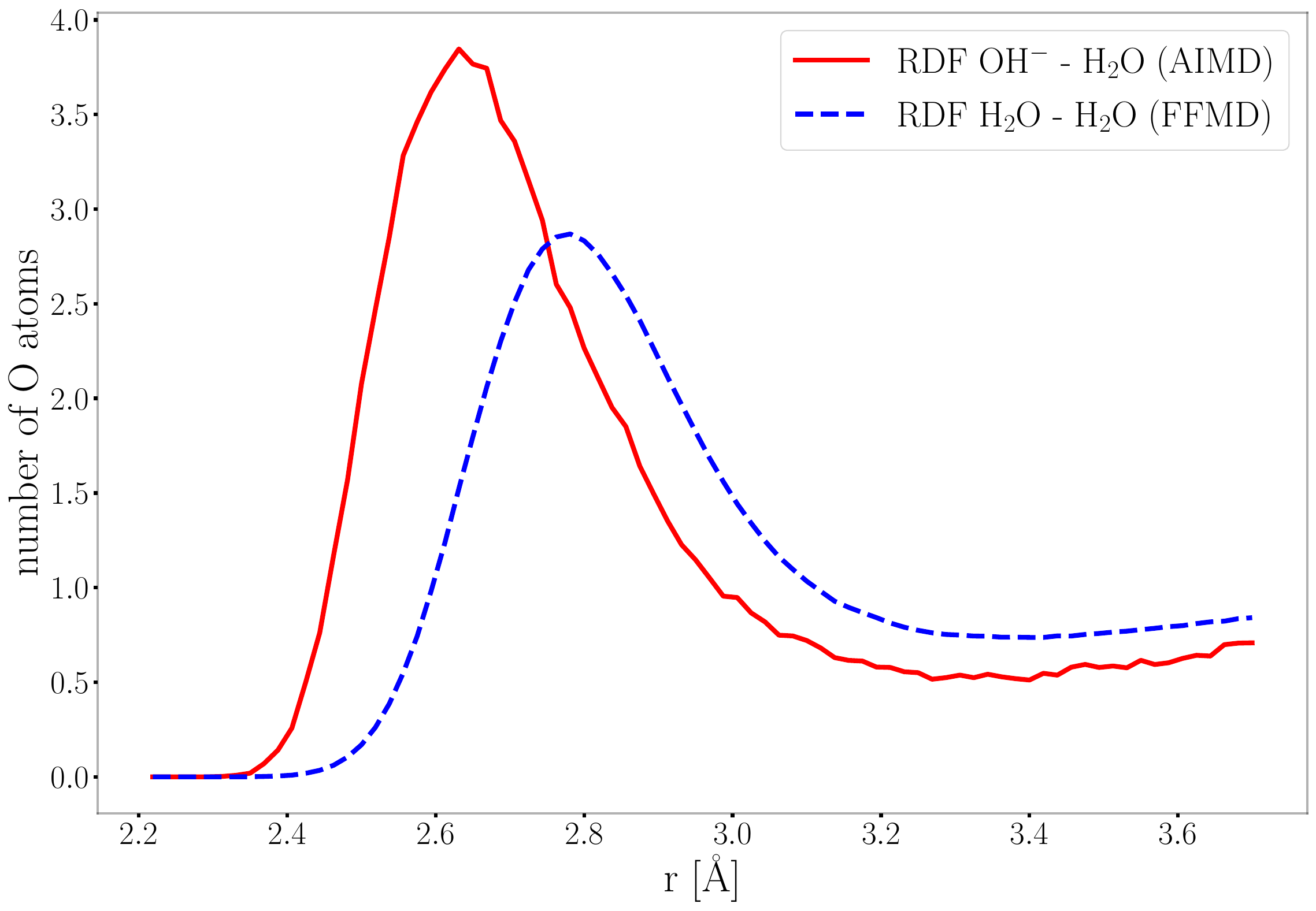}};
                    \begin{scope}[x={(image.south east)},y={(image.north west)}, overlay]
        	 			\node at  (0,0) {%
        	 					(a)
        	 			};
        	 		\end{scope}
	        \end{tikzpicture}
            \label{dist_rescaling1}
        \end{subfigure}%
        \vfill
    \end{minipage}
    \hfill
    \begin{minipage}[t]{0.35\textwidth} 
        \centering
        \vfill         
        \begin{subfigure}[t]{\textwidth} 
            \centering
            \begin{tikzpicture}
    	 		\centering
    	 		\node[anchor=north west,inner sep=0] (image) 
    	 		at (0,0) 
    	 		{\includegraphics[width=\textwidth]{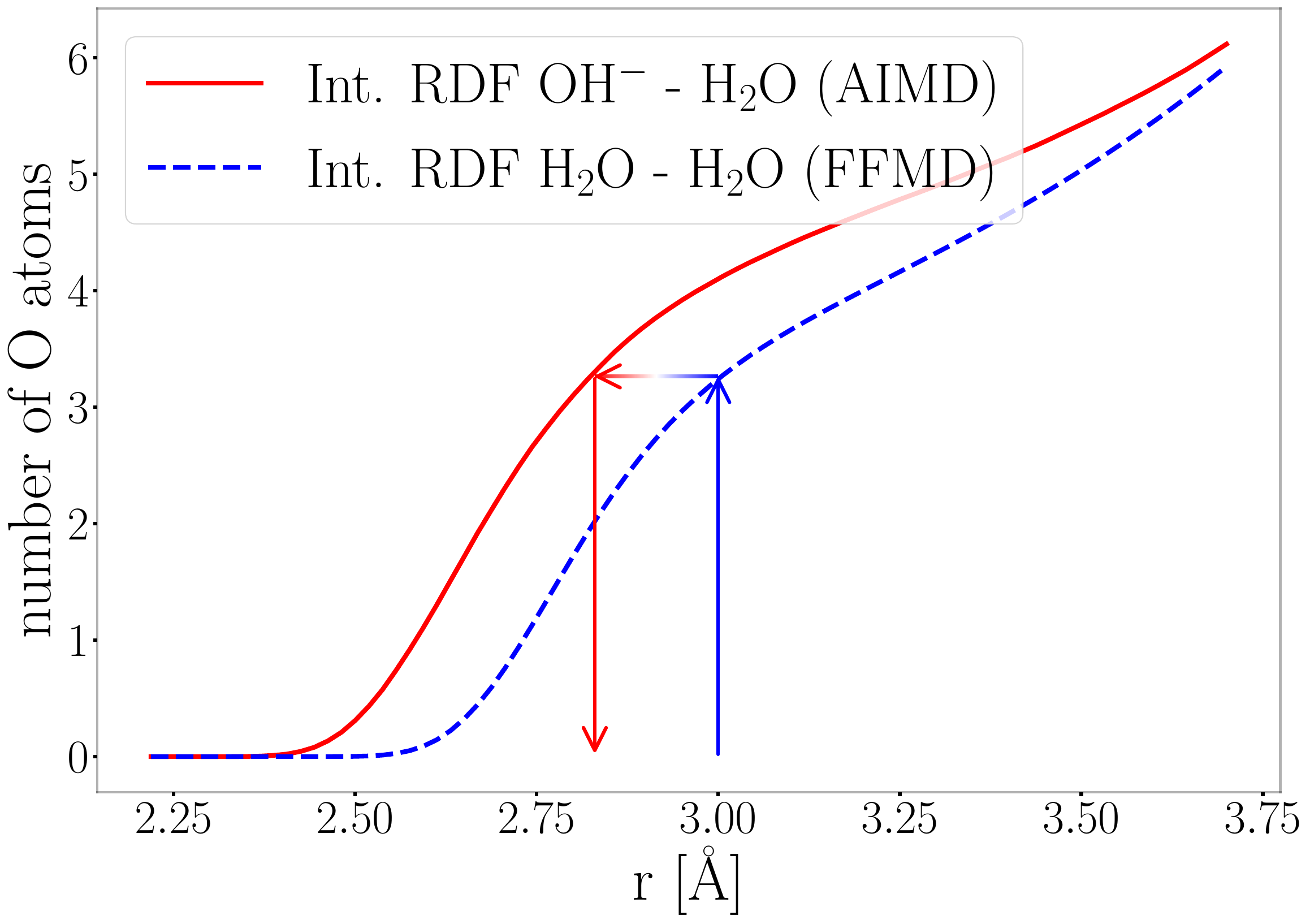}};
                \begin{scope}[x={(image.south east)},y={(image.north west)}, overlay]
        	 			\node at  (0,0) {%
                        		(b)
        	 			};
        	 		\end{scope}
	        \end{tikzpicture}
            \label{dist_rescaling2}
        \end{subfigure}
        \vspace{0.15cm}
        \begin{subfigure}[t]{\textwidth}
            \centering
            \begin{tikzpicture}
    	 		\centering
    	 		\node[anchor=north west,inner sep=0] (image) 
    	 		at (0,0) 
    	 		{\includegraphics[width=\textwidth]{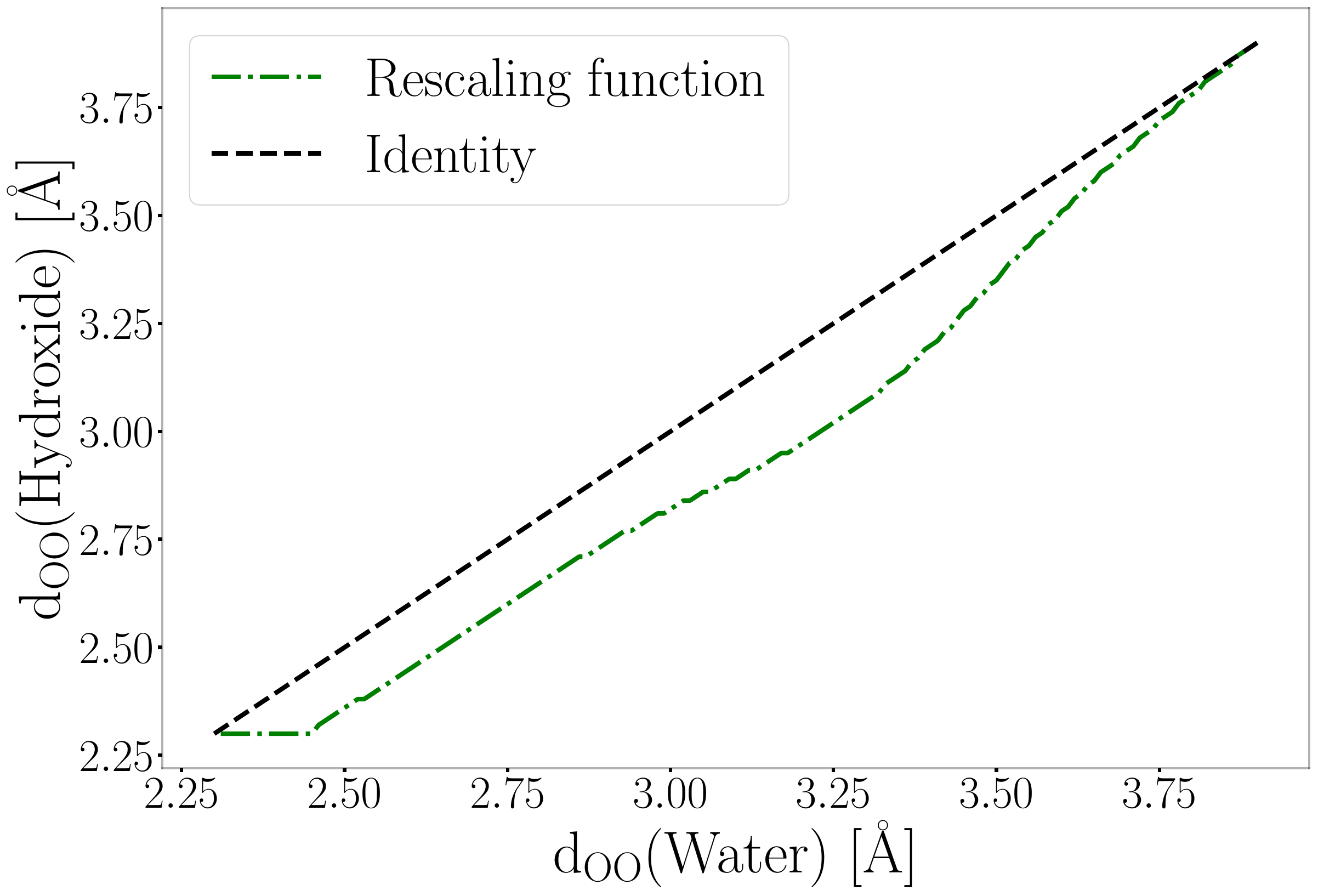}};
                    \begin{scope}[x={(image.south east)},y={(image.north west)}, overlay]
        	 			\node at  (0,0) {%
        	 					(c)
        	 			};
        	 		\end{scope}
	        \end{tikzpicture}
            \label{dist_rescaling3}
        \end{subfigure}
        \vfill
    \end{minipage}
    \caption{Distance rescaling process: (a) Radial distribution function (RDF) of the \ce{OO} distance for water and hydroxide ion, (b) the integrated OO-RDF and (c) the rescaling function, that maps water-water $d_\mathrm{OO}$ to hydroxide-water $d_\mathrm{OO}$.} 
    \label{img:dist_rescaling}
\end{figure}
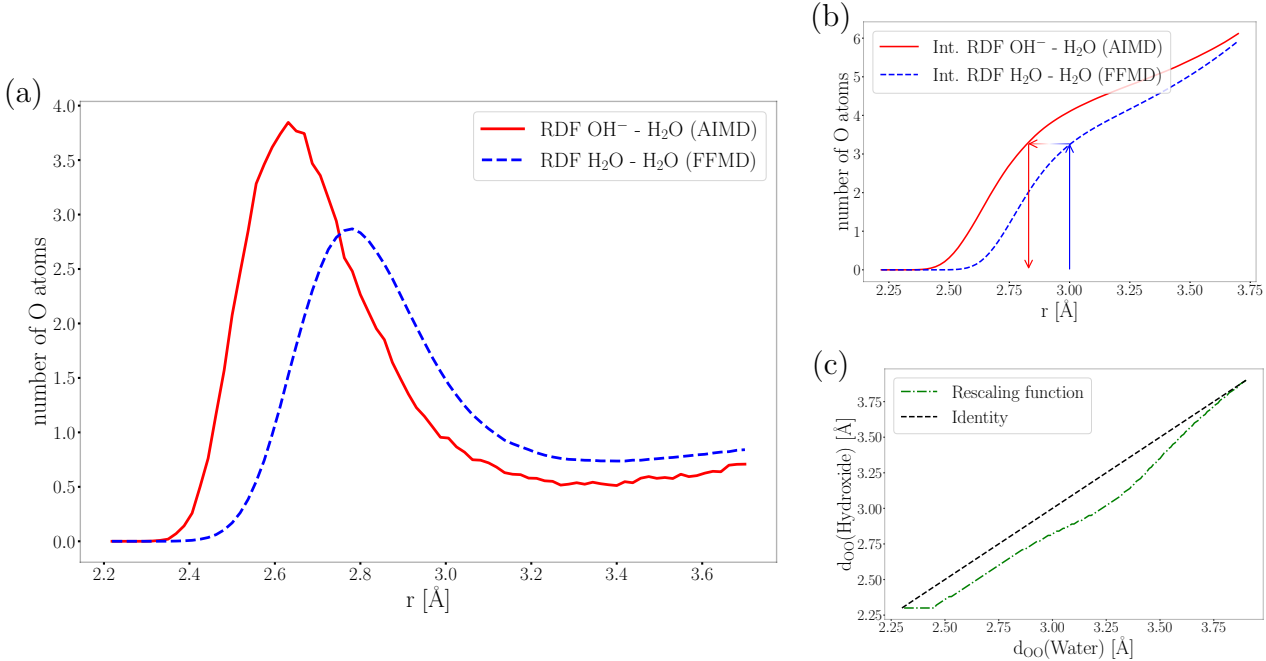

The average \ce{O-O} distance between a hydroxide ion and its nearest water molecule (\SI{2.6}{\angstrom}) is slightly shorter than the average distance between two neutral water molecules (\SI{2.75}{\angstrom}) (see Figure \ref{img:dist_rescaling}a). This discrepancy is evident in the integrated radial distribution functions (RDFs) shown in Figure \ref{img:dist_rescaling}b and poses a challenge for our proposed multiscale propagation scheme, because the topology of the Monte Carlo Lattice (on which the jump rates $\omega(d_{\mathrm{OO}})$ are applied) is constructed from a force field molecular dynamics simulation of pure water, i.e. in the absence of any \ce{K+} and \ce{OH-} ions. The application of the hopping rate function $\omega(d_\mathrm{OO})$, in turn, requires as input the oxygen-oxygen distance of a hydroxide-water pair (and not the distance between a water dimer). Within our multiscale scheme, the solution of this mismatch is an adequate transformation of the distance distribution which maps the $d_\mathrm{OO}$ distribution of water dimers onto the $d_\mathrm{OO}$ distribution of \ce{OH-}-\ce{H2O} by means of a rescaling function (see Figure \ref{img:dist_rescaling}c), previously employed in simulating long-range proton dynamics in acidic solutions.\cite{kabbe2017}

This function is derived from the integrated RDFs and ensures that the number of neighboring water molecules within a given radius r around a water molecule matches the number around a hydroxide ion. By aligning the structural environment of water and hydroxide ions, the rescaling function enables accurate application of the jump rate function in the multiscale framework.

\subsection{Simulation Scheme}
The combined Molecular Dynamics/Lattice Monte Carlo approach is implemented in the following stages:
\begin{enumerate}
\item AIMD Simulation: A brief \textit{ab initio} molecular dynamics simulation (\SI{20}{\pico\second}) of the aqueous potassium hydroxide solution is conducted to determine the jump rate function $\omega(d_{ij})$ based on Equation \ref{eq:fermi}.
\item Force field MD Simulation: An independent molecular dynamics simulation on nanosecond timescales is performed on pure water, free of ions, to generate the dynamic oxygen lattice. This simulation should be done such that it captures both short- and long-term structural fluctuations (e.g. hydrogen bond network correlations, local density fluctuations) which can be relevant to the diffusion mechanism in water. As a result, it provides an ideal lattice for the Monte Carlo approach. It should be noted that for a specific system (here: a given KOH solution), the resulting oxygen-oxygen distance distribution has to be further refined.
\item Hydroxide Ion Movement: Our Monte Carlo algorithm propagates protons on the oxygen lattice, using the transfer probabilities derived from the jump rate function (first stage) applied to the oxygen-oxygen (\ce{O-O}) distances (second stage). The \ce{O-O} distances are obtained from a trajectory of pure water and are therefore refined to be applicable to aqueous KOH systems. A rescaling function is used to adjust the \ce{O-O} distances before applying the jump rate function.
\end{enumerate}
The multiscale nature of our approach is evident in the significant computational efficiency achieved: the cost of propagating protons in the Monte Carlo step is reduced by several orders of magnitude compared to the cost of an equivalent AIMD step, enabling simulations at extended timescales with manageable computational resources.

\section{Results}

\begin{figure}[h!] 
    \centering 
    \includegraphics[width=1\textwidth]{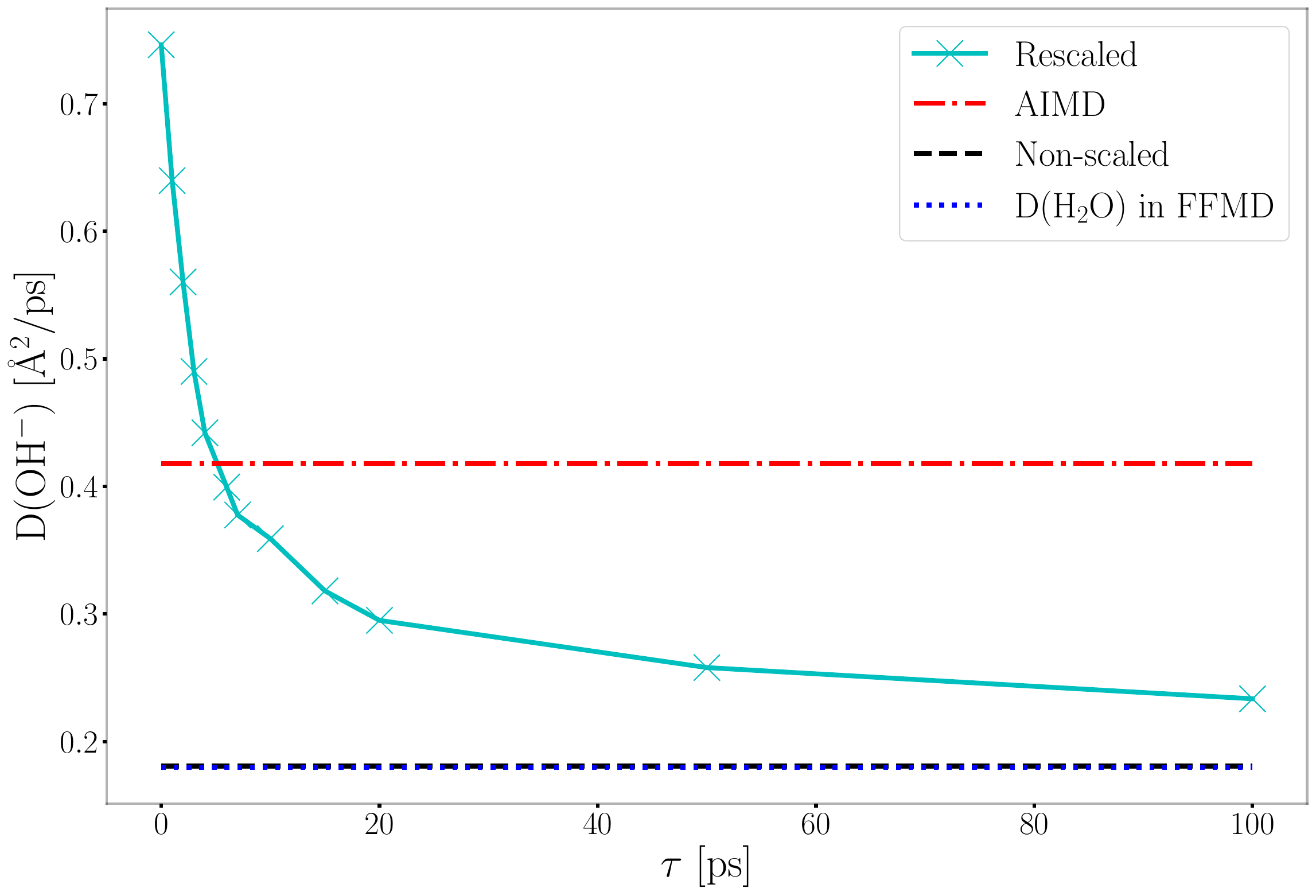} \caption{Hydroxide ion diffusion coefficient as a function of relaxation parameter $\tau$. The value $\tau$ = \SI{4}{\pico\second} provides the most accurate alignment with the AIMD simulation.} 
    \label{img:tau-dia} 
\end{figure}

We applied the proposed cMD/LMC approach to calculate the hydroxide diffusion coefficient, both with and without rescaling the \ce{O-O} distances. In the absence of rescaling, the diffusion coefficient of the hydroxide ions is significantly underestimated (\SI{0.18}{\square\angstrom\per\pico\second}) compared to the value obtained from AIMD simulations (\SI{0.42}{\square\angstrom\per\pico\second}) (see Figure \ref{img:tau-dia}). The predicted value from the approach without rescaling is comparable to the diffusion coefficient of water molecules observed in both AIMD and force field MD simulations. This is because proton jumps occur very rarely due to the larger distance between two water molecules, $d_\mathrm{OO}$ (unscaled distance values).

Instantaneous rescaling leads to an overestimation of the hydroxide diffusion coefficient (\SI{0.75}{\square\angstrom\per\pico\second}), a result also observed in the initial formulation of the multiscale approach for simulating proton dynamics in water.\cite{kabbe2017} Consequently, time-dependent rescaling is necessary, and a relaxation parameter ($\tau$) has been introduced to regulate the temporal evolution of \ce{O-O} distances. When $\tau$ = \SI{4}{\pico\second}, the diffusion coefficient aligns closely with the AIMD result (Figure \ref{img:tau-dia}), yielding  $D$\textsubscript{cMD/LMC}(\ch{OH-}) = \SI{0.44}{\square\angstrom\per\pico\second}.

To evaluate the diffusion coefficients of the hydroxide ions in the lengthy AIMD simulations (\SI{200}{\pico\second}) experimental values for ionic conductivity were used for comparison. 

\begin{equation}
    \sigma = \frac{D \cdot q^2 \cdot c(\mathrm{KOH}) \cdot N_A}{k_B \cdot T}
    \label{eq:cond}
\end{equation}

The computational ionic conductivities, derived from the AIMD simulations, were calculated using the equation \ref{eq:cond} where \( q \) is the charge of the moving ion. For monovalent ions, this corresponds to the \textit{elementary charge} (\SI{1.602e-19}{\coulomb}). \( D \) represents the diffusion coefficient, \( k_B \) is the \textit{Boltzmann constant} (\SI{1.381e-23}{\joule\per\kelvin}), \( T \) is the temperature, \( N_A \) is the \textit{Avogadro constant} (\SI{6.022e23}{\per\mole}), and \( c(\mathrm{KOH}) \) is the concentration of KOH in the solution.

The ionic conductivities computed from the AIMD simulations align with the experimental trend, confirming the reliability of AIMD simulations in capturing local hydroxide transfer rates to use in the cMD/LMC approach. The best agreement between AIMD-derived and experimental conductivity values is observed for dilute KOH solutions (\SI{0.56}{\mole\per\liter}: $\sigma$\textsuperscript{exp.} = \SI{0.13}{\siemens\per\centi\meter} and $\sigma$\textsuperscript{comp.} = \SI{0.12}{\siemens\per\centi\meter}) although the agreement slightly deteriorates at higher concentrations. 

\begin{figure}[h!]
    \centering
    \includegraphics[width=1\textwidth]{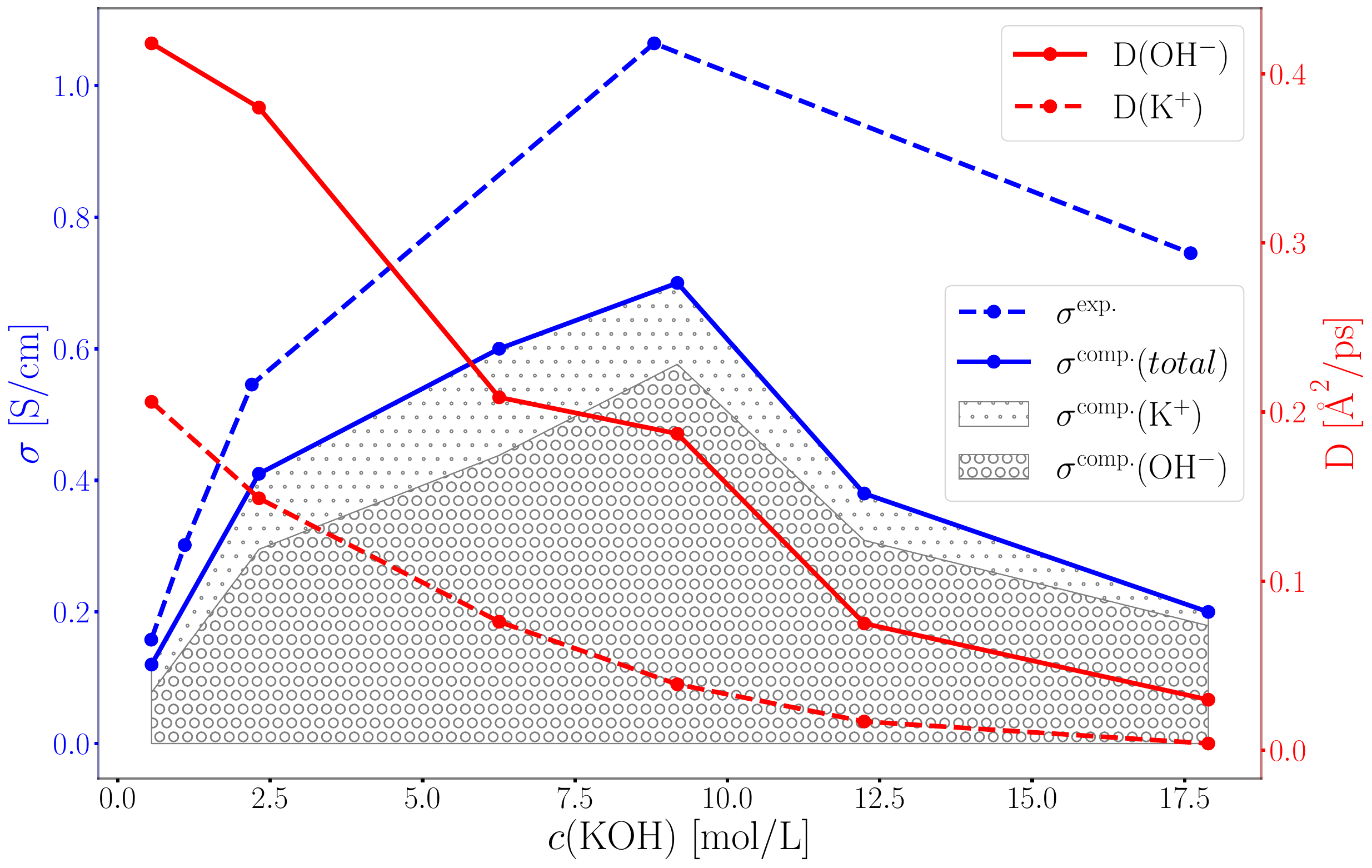}
    \caption{Ionic conductivity $\sigma$ and diffusion coefficients of \ce{OH-} and \ce{K+} ions as a function of KOH concentration at \SI{333}{\kelvin}.}
    \label{img:ionic_cond}
\end{figure}

Since the proton jump behavior in the cMD/LMC approach originates from AIMD, the comparison of $D$\textsubscript{AIMD}(\ch{OH-}) and $D$\textsubscript{cMD/LMC}(\ch{OH-}) remains the primary focus and will be addressed in detail after the next section.

\begin{figure}[h!]
    \centering
    \includegraphics[width=1\textwidth]{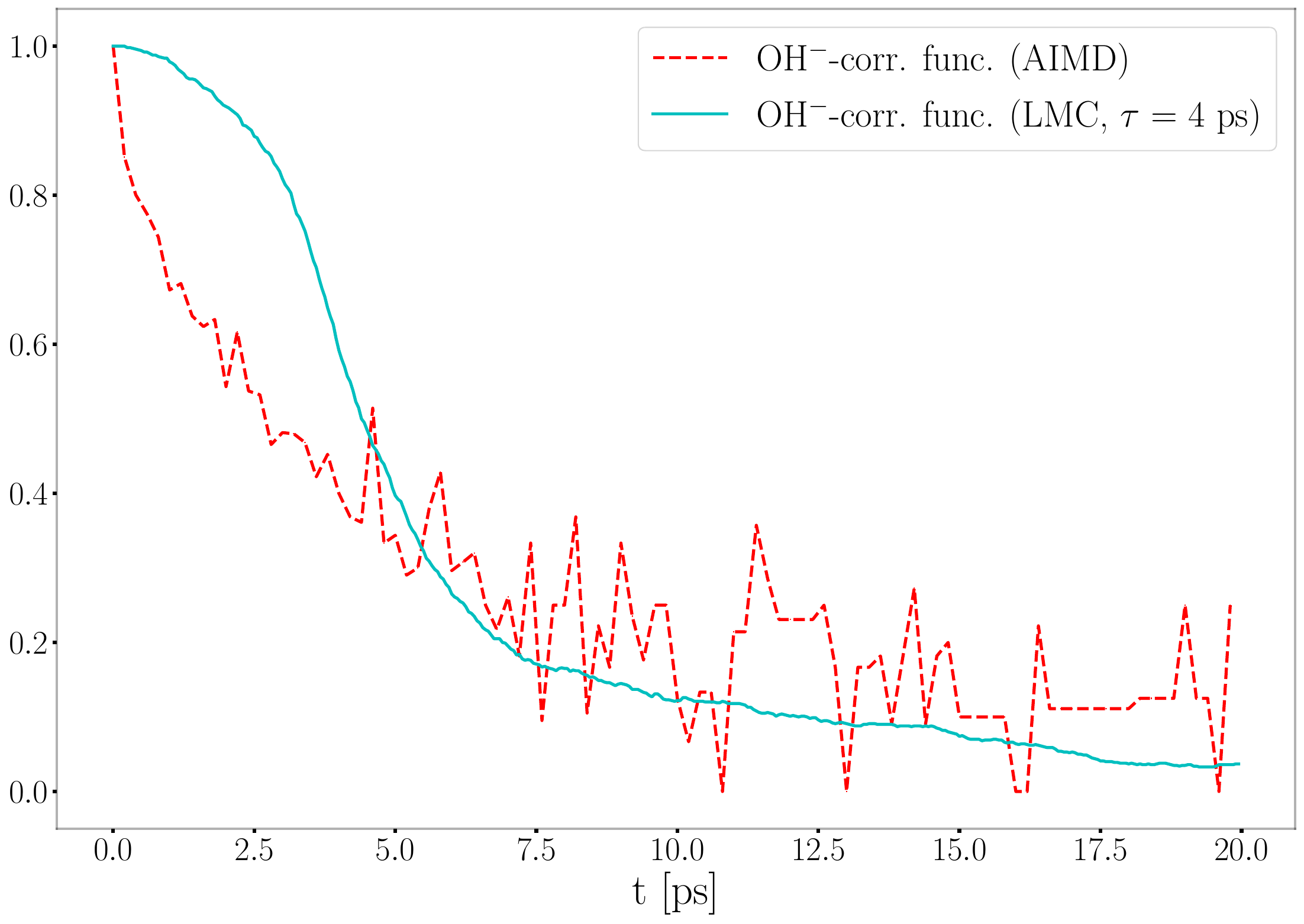}
    \caption{Hydroxide lifetime function obtained from the AIMD trajectory and the cMD/LMC approach.}
    \label{img:oh-lifetime}
\end{figure}

Another notable outcome of the cMD/LMC approach is the \ce{OH-} lifetime correlation function, which provides  insights into the average lifetime of hydroxide ions. In this context, lifetime refers to the duration of existence of a single protonated oxygen atom. A comparison between AIMD and cMD/LMC results shows that after an initial transient phase — during which the rescaling process in the cMD/LMC approach begins — both methods exhibit highly consistent behavior. The AIMD correlation function decays to zero at approximately \SI{15}{\pico\second}, while the cMD/LMC correlation function demonstrates a comparable decay profile, underscoring the accuracy of the method (see Figure \ref{img:oh-lifetime}). Furthermore, the half-life of the AIMD correlation function is \SI{3.5}{\pico\second}, closely matching the relaxation parameter $\tau$ employed in the cMD/LMC framework.

\section{Outlook}

\begin{figure}[h!]
    \centering
    \includegraphics[width=1\textwidth]{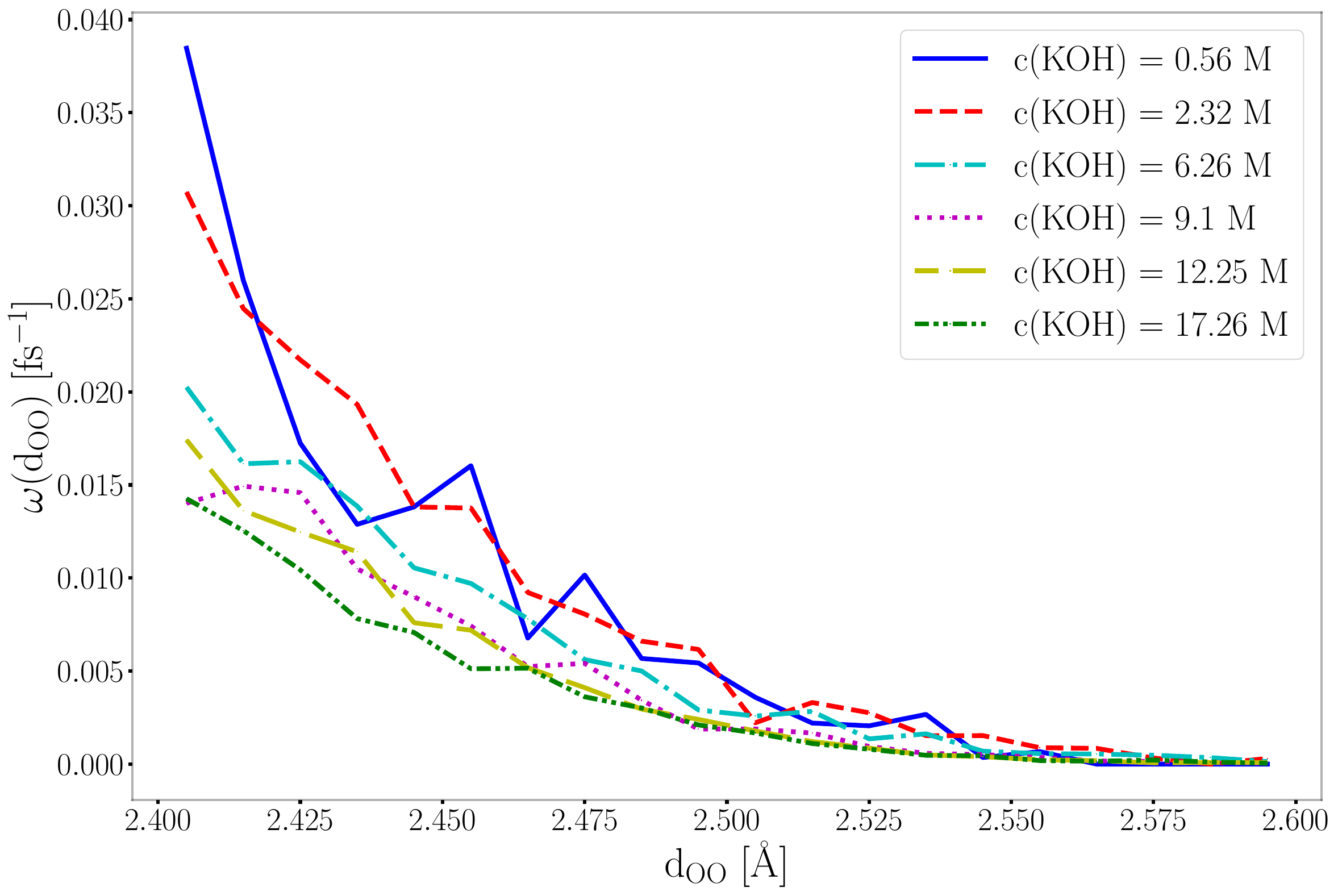}
    \caption{Jump rate functions of the AIMD trajectories with $c$(KOH) = \SI{0.56}{\mole\per\liter} to \SI{17.89}{\mole\per\liter}.}
    \label{img:all_jr}
\end{figure}

The efficiency of the algorithm is significantly enhanced by enabling the determination of the hydroxide ion diffusion coefficient at various concentrations of potassium hydroxide solutions using only a single short AIMD simulation.
A comparison of the jump rate functions obtained from AIMD simulations at different c(KOH) values shows no concentration dependence in the critical range of \SI{2.4}{\angstrom} to \SI{2.50}{\angstrom} (see Figure \ref{img:all_jr}). Consequently, the diffusion coefficients of hydroxide ions across different KOH concentrations can be qualitatively estimated with the cMD/LMC approach using a single jump rate function obtained from a short AIMD simulation.

\begin{figure}[h!]
    \centering
    \includegraphics[width=1\textwidth]{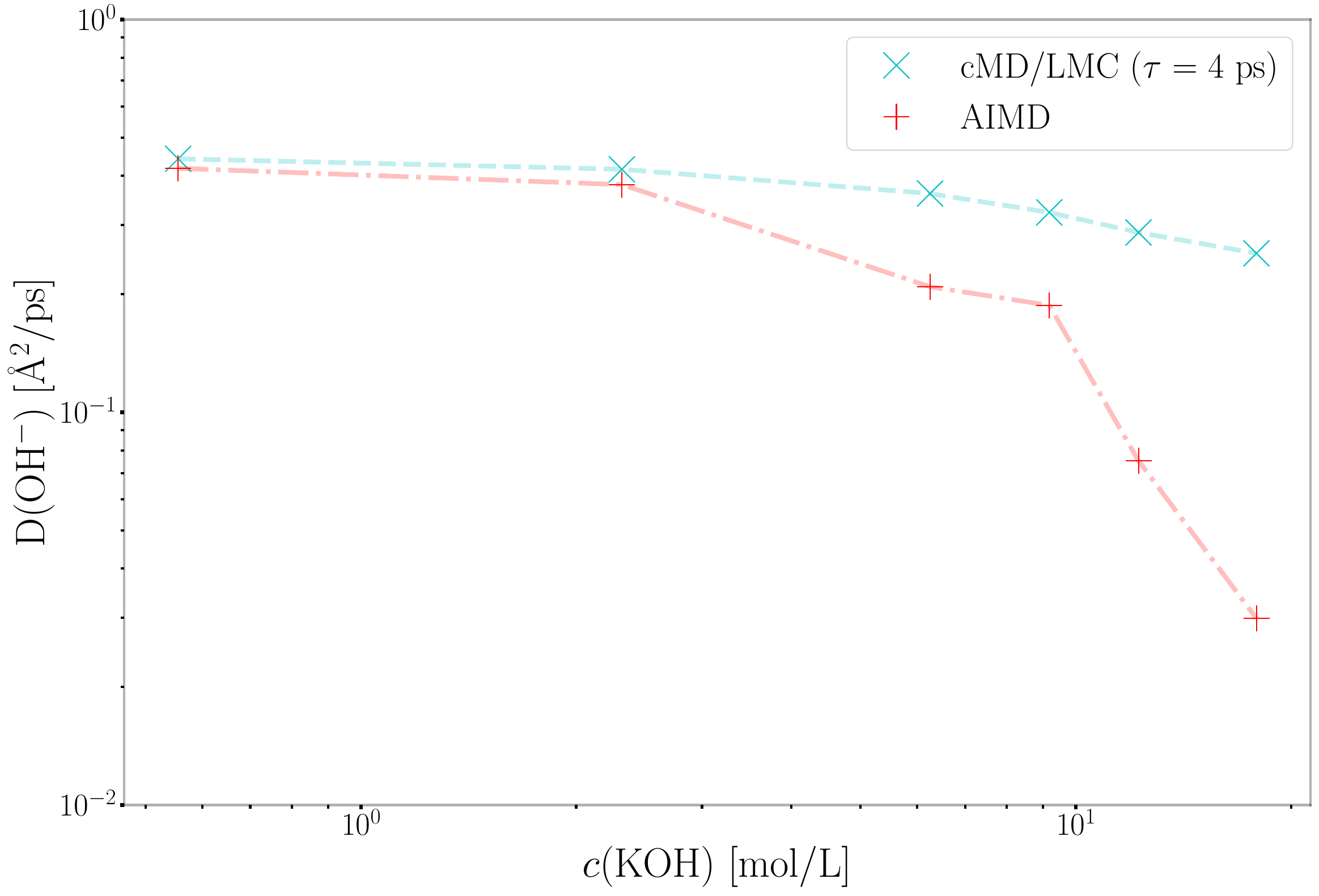}
    \caption{Diffusion coefficients of hydroxide ions at various KOH concentrations, obtained from AIMD simulations and the cMD/LMC method.}
    \label{img:diff_coeff}
\end{figure}

\begin{figure}[h!]
    \centering
    \includegraphics[width=1\textwidth]{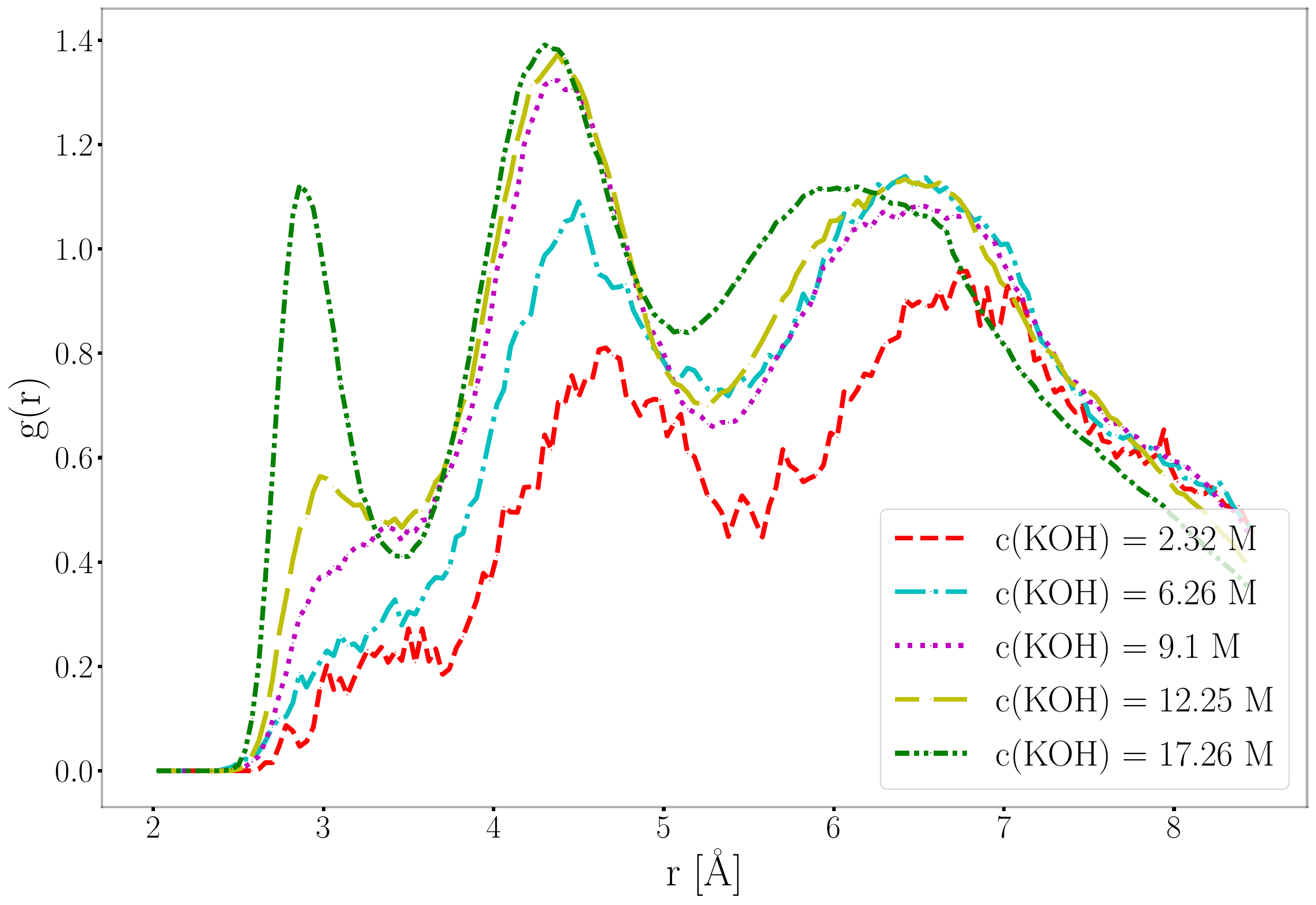}
    \caption{Radial distribution function between the oxygen atoms of two hydroxide moieties.}
    \label{img:rdf_oxox}
\end{figure}

The hydroxide ion diffusion coefficient at $c(\mathrm{KOH})$ = \SI{0.56}{\mole\per\liter} closely matches the value derived from AIMD simulations.
At $c$(KOH) = \SI{2.32}{\mole\per\liter}, the cMD/LMC method provides a reasonably accurate prediction of the diffusion coefficient (see Figure \ref{img:diff_coeff}). However, at higher KOH concentrations, the method's accuracy decreases due to increased ion interactions, which hinder diffusion, a trend that is reflected in the radial distribution function of the oxygen atoms of the hydroxide ions (see Figure \ref{img:diff_coeff} and \ref{img:rdf_oxox}). 
At potassium hydroxide concentrations of c(KOH) $\geq$ \SI{12.25}{\mole\per\liter}, a distinct first peak appears in the RDF, at a short distance of approximately \SI{2.9}{\angstrom} (see Figure \ref{img:rdf_oxox}). This feature is absent at lower concentrations. At these elevated concentrations, hydroxide ions displace water molecules from the solvation shells of other hydroxide ions, resulting in a separation of approximately \SI{2.9}{\angstrom} between hydroxide oxygen atoms. This proximity enables the formation of weak hydrogen bonds despite the repulsive electrostatic interactions.

Despite this limitation, the method remains effective for capturing the general trend of hydroxide ion diffusion across a range of concentrations.
As the KOH concentration increases, the viscosity of the solution also increases. To account for this, we adjusted the viscosity according to experimental values by changing the temperature in the simulation. Details of the modified force field MD simulations are provided in the computational details section.

\FloatBarrier

Another study, published in December 2024, focuses on predicting the diffusion of hydroxide ions in an aqueous environment using a similar method.\cite{dutta2024} This approach is also based on FFMD simulations combined with a mechanism to model proton jumps. Unlike our method, the model proposed by Dutta et al. alternates between Monte Carlo and molecular dynamics (MD) steps, allowing proton jumps to influence the subsequent trajectory of the oxygen atoms.

In contrast, our approach involves generating the water trajectory using FFMD simulations and then propagating protons on the lattice of the oxygen atoms via a Monte Carlo method. Furthermore, unlike Dutta et al., the jump probabilities in our Monte Carlo steps are derived from a distance-dependent jump rate function, which we determine through a short AIMD simulation of aqueous hydroxide solution. The method described in the aforementioned study utilizes a modified Metropolis criterion with an empirical threshold parameter, calibrated to achieve the correct diffusion coefficient. This approach incorporates the coordination geometry of the hydroxide ion’s first solvation shell.

Since our method relies on a pre-generated water trajectory (without hydroxide ions), we employ a distance-rescaling mechanism and a relaxation parameter to accurately represent the hydrogen-bond network dynamics in the presence of hydroxide ions. Future research could explore combining the method of Dutta et al. with our approach. Such a hybrid method might involve alternating MD and Monte Carlo steps (as in the approach of Dutta et al.) while utilizing a distance-dependent jump rate function, sampled from AIMD simulations, to calculate proton transfer probabilities (as in our approach).

\section{Conclusion}

Building on the computationally demanding AIMD method, which accurately simulates atomistic processes such as covalent bond formation and breakage (including proton diffusion via the Grotthuss mechanism) but is limited to a few hundred atoms over a few picoseconds, we adapted and applied a combined Molecular Dynamics/Lattice Monte Carlo approach. This method enables the modeling of hydroxide ion transfer over significantly larger timescales and extended system sizes. The proton transfer algorithm was successfully applied to a range of concentrations of aqueous KOH solutions.

The cMD/LMC approach delivers reliable results for KOH solutions at low concentrations.

However, at exceptionally high concentrations, the method's accuracy decreases due to 1) the elevated viscosity of the solutions, which is only incorporated in our simulation scheme by reducing the temperature in the underlying force field MD simulations and 2) the unusual formation of hydrogen bonds between hydroxid ions at very high hydroxide concentrations.

The empirical relaxation parameter $\tau$, used to describe the hydrogen-bond network's response time to a proton jump, was set to \SI{4}{\pico\second}. 
This value aligns closely with the mean time between proton jumps in the AIMD simulation (\SI{3}{\pico\second}) and the half-life of the hydroxide lifetime correlation function (\SI{3.5}{\pico\second}).

The cMD/LMC approach provides a computationally efficient framework for simulating aqueous KOH solutions, covering timescales from nanoseconds to milliseconds and systems involving several thousand atoms. This method enables the detailed characterization of \ce{OH-} dynamics in complex nanostructered systems, including anion-exchange membranes and layered double hydroxides, offering a valuable tool for advancing research in these areas.

\section{Computational Details}
\subsection{\textit{ab initio} molecular dynamics simulation of KOH in aqueous solution}
Systems with KOH concentrations from \SI{0.56}{\mole\per\liter} to \SI{17.89}{\mole\per\liter} were simulated. Their details are listed in table \ref{tab:aimd_KOH}.
\begin{table}[h!]
    \centering
    \caption{Computational details of the AIMD simulations.}
        \begin{tabular}{l l l l}
        \toprule
        simulation of   &	1 KOH              &  4 KOH            & 10 KOH         \\
        			    &	in 98 \ch{H2O}     &  in 92 \ch{H2O}   & in 80 \ch{H2O} \\
        \midrule
        $c$(KOH) [\si{\mole\per\liter}]   & 0.56 & 2.32 & 6.26 \\
        w(KOH) [\si{\percent}]          & 3 & 12 & 28 \\ 
        box size:   &&& \\
        \hspace{1.8cm}$x$ [\si{\angstrom}]    & 14.41  & 14.21 & 13.85  \\
        \hspace{1.8cm}$y$ [\si{\angstrom}]    & 14.41  & 14.21 & 13.85  \\
        \hspace{1.8cm}$z$ [\si{\angstrom}]    & 14.41  & 14.21 & 13.85  \\
        angle [\si{\degree}]    & $\alpha=90$ & $\beta=90$ & $\gamma=90$ \\
        number of atoms	        & 297& 288 & 270 \\
        duration of time step [\si{\femto\second}]  	& 0.5 & 0.5 & 0.5 \\
        temperature	[\si{\kelvin}]	& 333  & 333 & 333 \\
        simulation time [\si{\pico\second}]	& 200 & 200 & 200  \\
        energy drift [\si{Ha\per\femto\second}]	&$4.7\cdot10^{-9}$		&	$9.8\cdot10^{-8}$ & $1.6\cdot10^{-7}$ 	 \\	
        \bottomrule					
               \toprule
        simulation of   &	14 KOH              &	18 KOH 	          &  25 KOH                            \\
        			    &	in 72 \ch{H2O}     &    in 64 \ch{H2O} 	  &  in 50 \ch{H2O}        \\
        \midrule
        $c$(KOH) [\si{\mole\per\liter}]   & 9.18 & 12.25 & 17.89   \\
        w(KOH) [\si{\percent}]          & 37 & 48 & 61    \\ 
        box size:   &&& \\
        \hspace{1.8cm}$x$ [\si{\angstrom}]    & 13.64 & 13.46 & 13.24    \\
        \hspace{1.8cm}$y$ [\si{\angstrom}]    & 13.64 & 13.46 & 13.24   \\
        \hspace{1.8cm}$z$ [\si{\angstrom}]    & 13.64 & 13.46 & 13.24   \\
        angle [\si{\degree}]    & $\alpha=90$ & $\beta=90$ & $\gamma=90$  \\
        number of atoms	        & 258 & 246 & 225   \\
        duration of time step [\si{\femto\second}]  	& 0.5 & 0.5 & 0.5  \\
        temperature	[\si{\kelvin}]	& 333  & 333 & 333   \\
        simulation time [\si{\pico\second}]	& 250 & 250 & 350 \\
        energy drift [\si{Ha\per\femto\second}]	& $5.7\cdot10^{-8}$ & $1.1\cdot10^{-7}$ & $1.1\cdot10^{-8}$  \\	
        \bottomrule		
    \end{tabular}
    \label{tab:aimd_KOH}
\end{table}
\par
The structures were subjected to a geometry optimization before simulation. The software package CP2K\cite{cp2k_1, cp2k_2, cp2k_3} for quantum chemistry and solid state physics was used for this purpose as well as for the \textit{ab initio} molecular dynamic simulations that were performed at temperatures of \SI{333}{\kelvin}. The trajectories comprise \SI{200}{\pico\second} with a timestep every \SI{0.5}{\femto\second}. 
\par
The electronic structure was modeled with these quantum chemical calculations utilizing the density-functional theory (DFT).\cite{DFT1-Theorem_Hohenberg,DFT2_Kohn_Sham1,DFT3_Kohn_Sham2} The module Quickstep\cite{cp2k_quickstep} and an efficient orbital transformation method\cite{cp2k_orb_trans} were chosen in favor of a fast convergence. The BLYP-functional was used as the XC-functional.\cite{cp2k_blyp1,cp2k_blyp2} Moreover a basis set of the type DZVP-MOLOPT-SR-GTH\cite{cp2k_basis-set} \!, GTH-BLYP pseudo-potentials \cite{cp2k_gth-pseudopot1,cp2k_gth-pseudopot2, cp2k_gth-pseudopot3} and the empirical dispersion correction form Grimme (D3)\cite{cp2k_d3, cp2k_d3_2} were applied. The jumps which are part of the long-range proton transfer from these simulations are defined to be real jumps.

\subsection{Force field molecular dynamics simulation of water}
The classical molecular dynamics simulations were obtained with the "Large-scale Atomic/Molecular Massively Parallel Simulator" (LAMMPS) utilizing the TIP4P water model.\cite{LAMMPS, horn_tip4p} Their details are listed in Table \ref{tab:cmd_H2O}. The temperature of the FFMD simulation was adjusted to ensure that the diffusion coefficients of water molecules in FFMD matched the values obtained from AIMD at higher concentrations.
\begin{table}[h!]
    \centering
    \caption{Computational details of the FFMD simulations.}
    \begin{tabular}{l l l l l l l}
        \toprule
        simulation of               &	256 \ch{H2O}  &   256 \ch{H2O}    &   256 \ch{H2O}    &   256 \ch{H2O}    &   256 \ch{H2O}    &   256 \ch{H2O} \\
        emulating w(KOH) [\si{\percent}] &  3 & 12 & 28 & 37 & 48 & 61 \\
        \midrule
        temperature	[\si{\kelvin}]	& 288  & 278 & 268 & 253 & 243 & 228\\
        box size:   &&&&& \\
        \hspace{1.8cm}$x$ [\si{\angstrom}]    & 19.71 & 19.71 & 19.71 & 19.71 & 19.71 & 19.71 \\
        \hspace{1.8cm}$y$ [\si{\angstrom}]    & 19.71 & 19.71 & 19.71 & 19.71 & 19.71 & 19.71 \\
        \hspace{1.8cm}$z$ [\si{\angstrom}]    & 19.71 & 19.71 & 19.71 & 19.71 & 19.71 & 19.71 \\
        angle [\si{\degree}]    & \multicolumn{6}{c}{$\alpha=90$; \hspace{2.5cm}$\beta=90$; \hspace{2.5cm} $\gamma=90$} \\
        number of atoms	        & 768 & 768 & 768 & 768 & 768 & 768 \\
        duration of time step [\si{\femto\second}]  	& 0.5 & 0.5 & 0.5 & 0.5 & 0.5 & 0.5 \\
        simulation time [\si{\pico\second}]	& 2500 & 2500 & 2500 & 2500 & 2500 & 2500 \\
        \bottomrule					
    \end{tabular}
    \label{tab:cmd_H2O}
\end{table}

\FloatBarrier
\section{Experimental Details}

\begin{figure}[h!]
    \centering
    \includegraphics[width=0.6\textwidth]{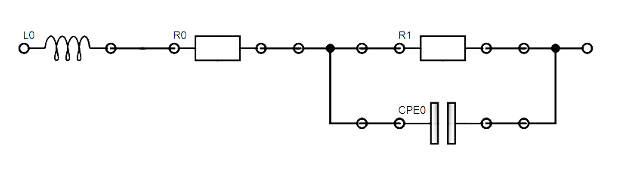}
    \caption{Equivalent circuit for fitting EIS measurements consisting of an inductor, two ohmic resistors and a constant phase element (CPE).}
    \label{img:cond_exp}
\end{figure}

The conductivity of the KOH solution was measured via electrochemical impedance spectroscopy (EIS) with a potentiostat (Zahner\textsuperscript{\textregistered} Zennium Pro) connected to an in-house made cell that was heated in an oven. Two symmetrical round metal plates made from nickel (Ni 2.4060) were used as electrodes. The surface area wetted by the electrolyte was \SI{18.2}{\square\centi\meter} and the distance between the electrodes was \SI{3.17}{\centi\meter}. The electrolyte was freshly prepared from KOH pellets ($\geq$ \SI{85.0}{\percent}) with Millipore water (\SI{18.2}{\mega\ohm\centi\meter} at \SI{25}{\celsius}). EIS was carried out potentiostatic at \SI{100}{\milli\volt} with an amplitude of \SI{10}{\milli\volt} and a frequency range of 1 to \SI{300000}{\hertz}. 4 steps per decade and 4 measure periods were chosen below \SI{66}{\hertz} and 10 steps per decade and 20 measure periods above \SI{66}{\hertz}. The measurement result was fitted to an equivalent circuit (see Figure \ref{img:cond_exp}) consisting of an inductor, two ohmic resistors and a constant phase element (CPE). 

$R_0$ was used to calculate the conductivity of the KOH solution $\sigma^{\textrm{exp.}}$ with equation \ref{eq:exp_cond}. $A$ is the surface area wetted by the electrolyte and $l$ is the distance between the electrodes.
\begin{equation}
    \sigma^{\textrm{exp.}} = R_0 \frac{A}{l}
    \label{eq:exp_cond}
\end{equation}
\FloatBarrier

\begin{suppinfo}

The Supporting Information is available from the Wiley Online Library or from
the author.

\end{suppinfo}

\begin{acknowledgement}

We gratefully acknowledge financial support by TAB research group "KapMemLyse".
We also thank the staff of the Compute Center of the Technische Universität Ilmenau and especially Mr.~Henning~Schwanbeck for providing an excellent research environment. 

\end{acknowledgement}

\section{Conflict of Interest}
The authors declare no conflict of interest.

\section{Data Availability Statement}
The data that support the ﬁndings of this study are available from the cor-
responding author upon reasonable request.

\section{Keywords}
anion-exchange membrane water electrolysis, proton transport, hydroxide ion dynamics, molecular dynamics simulations, multiscale simulations

\bibliography{bibliography.bib}

\end{document}